# Broad-band robustly single-mode hollow-core PCF by resonant filtering of higher order modes

**Mehmet C. Günendi, Patrick Uebel\*, Michael H. Frosz and Philip St.J. Russell**

*Max Planck Institute for the Science of Light, Guenther-Scharowsky-Str. 1, 91058 Erlangen, Germany*
[*patrick.uebel@mpl.mpg.de](mailto:patrick.uebel@mpl.mpg.de)

**Abstract:** We propose and theoretically analyse a novel hollow-core photonic crystal fibre (PCF) that is engineered so as to strongly suppress higher order modes, i.e., to provide robust $LP_{01}$ single-mode guidance in all the wavelength ranges where the fibre guides with low loss. Encircling the core is a single ring of non-touching glass elements whose modes are tailored to ensure resonant phase-matched coupling to higher-order core modes, causing them to leak at a very high rate into the supporting solid glass sheath. Using a model based on coupled capillary waveguides, as well as full vectorial finite element modelling, we show that this modal filtering effect depends on only one dimensionless geometrical parameter, akin to the well-known *d*/*Λ* parameter for endlessly single-mode solid-core PCF. The design is scalable up to large core sizes and is predicted to deliver $LP_{01}$ mode losses of some 10s of dB/km in multiple transmission windows, the broadest of which spans more than an octave. At the same time, higher order core modes have losses that are typically 1000s of times higher than the $LP_{01}$ mode. The fibre is of great potential interest in applications such as high-power laser beam delivery, gas-based nonlinear optics, laser particle delivery, chemical sensing and laser gyroscopes.

## 1. Introduction

Through their ability to guide light in low index materials such as gases and liquids, hollow-core photonic crystal fibres (HC-PCFs) provide many opportunities beyond those of solid-core fibres and are finding applications such as high speed "low-latency" data transmission [1], high power beam delivery [2, 3], gas-based nonlinear optics [4, 5] and chemical and gas sensing [6, 7]. HC-PCFs typically come in two classes according to guidance mechanism, i.e., whether it is via a two-dimensional photonic bandgap or anti-resonant-reflection (ARR).

Several different types of hollow core ARR-PCFs have emerged in recent years, including fibres with kagomé-style claddings [8] and a new class of simpler structure consisting of a single ring of anti-resonant elements (AREs) surrounding a central hollow core [9-11]. These single-ring structures are potentially easier to fabricate than kagomé-PCF and offer in some cases lower transmission losses, for example in the mid-IR [12]. Numerical studies suggest that they might even attain attenuation levels below those of the best solid-core telecommunication fibres, though this has not yet been proven experimentally [13].

Compared to solid-core fibres, in particular endlessly single-mode (ESM) PCF [14], a drawback of ARR-PCFs is that they also support a family of higher-order modes (HOMs), i.e., they are not purely single-mode waveguides. Since these HOMs often have relatively low loss, it is very difficult to launch a pure $LP_{01}$ mode without HOM contamination and - even if one succeeds in doing so - HOMs can be excited by bending or external stress, resulting in lower-quality transmitted beam profiles and power fluctuations. This problem is particularly acute in applications using relatively short fibre lengths, such as laser machining and pulse compression.

To quantify the degree of HOM suppression, we introduce a figure of merit

$$\text{FOM}_{ij} = \frac{\alpha_{ij} - \alpha_{01}}{\alpha_{01}} \tag{1}$$

where $\alpha_{ij}$ is the loss of the $LP_{ij}$ core mode (in dB/m). Applying this FOM to published results, we find that a structure with a single ring of AREs, each consisting of a pair of nested capillaries [13], offers $FOM_{11}$ values of order 600 over limited wavelength ranges, while a single close-packed ring of capillaries [15] offers $FOM_{11}$ values of 50 at a single wavelength. None of these approaches are, however, free of disadvantages: they are either very difficult to make, offer relatively high $LP_{01}$ mode loss or do not allow robust single-mode guidance in all wavelength regions where the $LP_{01}$ mode is supported.

Here, for the first time, we report a simple single-ring ARR-PCF design that is scalable and provides very strong HOM suppression within all the low loss guidance bands of the fibre. The design is based on a generalization of ESM guidance in solid-core PCFs [14]. As a convenient short-hand, we denote this new fibre type as hESM (hollow-core endlessly single-mode). Our focus is on achieving the highest possible suppression of HOMs while maintaining reasonably low loss for the $LP_{01}$ mode. Note that for clarity we will refer to the core modes using the $LP_{lm}$ notation and the modes of the surrounding capillaries using the notation $ARE_{lm}$, where $l$ and $m$ represent the azimuthal and radial orders of the modes.

The hESM structure consists of a central hollow core (inner diameter $D$) surrounded by six evenly spaced and non-touching capillaries (AREs) with wall thickness $t$ and inner diameter $d$ (see top panel in Fig. 1a), supported within a thick-walled supporting capillary. Our initial idea was that, within certain wavelength ranges, the AREs would provide anti-resonant reflection for the $LP_{01}$ mode, while the gaps between guaranteed leakage for the smaller-lobed higher order modes, in a similar manner to "modal sieving" in solid-core PCFs [16]. Although the reality turns out to be more complex and subtle, we have discovered by finite element (FE) modelling that extremely high values of $FOM_{11}$ (several thousand) can be achieved over all wavelength bands (some of which are octave-spanning) where the fibre guides with low loss, provided the ratio $d/D \approx 0.68$ and the ARE walls are thin enough. We provide a physical interpretation for the numerical results using a simple analytical model based on coupled capillaries.

## 2. Numerical design of hESM PCF

The modal index and loss for the $LP_{01}$ and $LP_{11}$ core modes of a straight hESM (modal field distribution plotted in the lower panels of Fig. 1a), calculated using FE modelling, are plotted in Fig. 1b against the ratio $d/D$ for $t/D = 0.01$ and $D/\lambda = 20$. The glass refractive index $n_g$ was set to a constant value of 1.45 (corresponding to silica at 1 µm) and the hollow regions were taken to be vacuum. For reference, the index and loss of the $ARE_{01}$ mode of an isolated capillary in vacuum are also included (brown dashed lines). Fig. 1b (upper) shows that the index of the $LP_{01}$ core mode is high enough to avoid resonant coupling to the $ARE_{01}$ modes, and remains almost independent of $d/D$, whereas the $LP_{11}$ core mode undergoes a strong anti-crossing with the $ARE_{01}$ mode at $d/D \approx 0.68$. As one moves away from this anti-crossing, the even and odd eigenmodes evolve asymptotically into uncoupled $LP_{11}$ and $ARE_{01}$ modes.

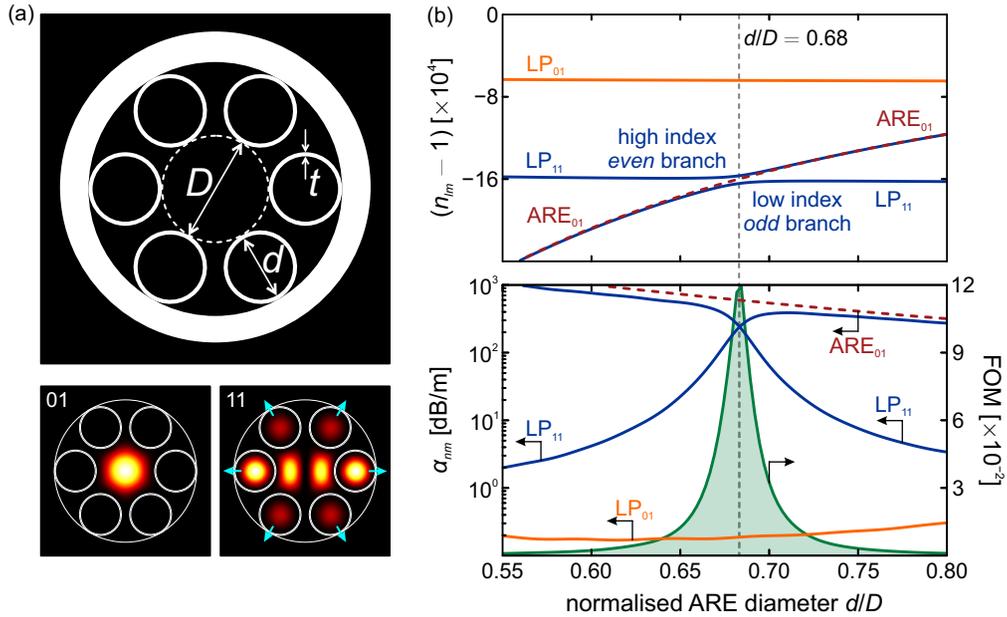

Fig. 1. (a) Top: Sketch of the hESM cross-section. Glass is marked in white and the hollow regions in black. Lower two panels: Modulus squared of the dominant electric field component (linear colour scale), calculated using FE modelling, for the $LP_{01}$ mode (left) and a hybrid $LP_{11}/ARE_{01}$ mode at the anti-crossing (right). The blue arrows indicate the main leakage channels for the selected hybrid mode (note that additional degenerate forms of this mode exist). (b) Numerically calculated dependence of the modal properties on $d/D$, keeping $t/D = 0.01$ and $D/\lambda = 20$. Upper: modal refractive indices of the $LP_{01}$ mode and the two hybrid $LP_{11}/ARE_{01}$ modes. Lower: modal losses and $FOM_{11}$. The brown dashed curves in each plot refer to the $ARE_{01}$ mode of an isolated ARE capillary.

Core modes of even higher order (not shown in Fig. 1) have lower indices and couple to highly leaky modes of the ARE ring, some of which are concentrated in the gaps between the AREs.

Fig. 1b (lower) plots the calculated leakage loss of the $LP_{01}$ mode and the two hybrid $LP_{11}/ARE_{01}$ modes. Over the range shown the $LP_{01}$ core mode has a relatively constant loss with a minimum value of 0.17 dB/m at $d/D \approx 0.65$ (for this particular scaling). For smaller ARE diameters the loss increases, closely matching the value for an isolated thick-walled dielectric capillary in the limit $d/D \to 0$ (not shown in Fig. 1b). This limit was used to cross-check the FE calculations with analytical results [17], in particular the accuracy of the perfectly matched layers (PMLs). At the anti-crossing point the loss of the two $LP_{11}/ARE_{01}$ modes strongly increases, almost reaching the value for an isolated capillary in vacuum (dashed brown line), which was calculated by solving Maxwell's equations in full vectorial form. This provides further confirmation that the PMLs were set up correctly.

The $FOM_{11}$ increases strongly at the anti-crossing (green curve in Fig. 1b), peaking at a value of ~1200. Far from the anti-crossing it drops to less than 5, which is similar to values typically achieved in kagomé-PCF [18]. For a comprehensive analysis of HOM suppression, the FOM of all the higher-order core modes must be calculated. FE modelling reveals that the HOM with the next-lowest loss after the $LP_{11}$ mode is the four-lobed $LP_{21}$ mode, with a $FOM_{21}$ of ~70 at $d/D \approx 0.68$ and an anti-crossing with the $ARE_{01}$ mode at $d/D \approx 0.53$. In experiments, however, this particular core mode is less likely to be excited by end-fire illumination or by stress- and bend-induced scattering from the $LP_{01}$ mode (the index difference is some two times larger than for the $LP_{11}$ mode). FE modelling shows that the $FOM_{lm}$ increases strongly for $LP_{lm}$ modes of even higher order, with the result that they have an insignificant impact on the performance of the hESM fibre. This is because they are phase-matched and strongly coupled to modes of the ARE ring (some of which are concentrated in the gaps between the AREs), resulting in high leakage loss.

To illustrate the scalability of the hESM structure we plot in Fig. 2a, versus $D/\lambda$, the effective refractive index $n_{lm}$ of the $LP_{lm}$ core modes and the $ARE_{01}$ mode at constant $d/D = 0.68$ and $t/D = 0.01$. $n_{01}$ increases with increasing $D/\lambda$ but overall remains above the index of the $ARE_{01}$ mode. As a consequence, the $LP_{01}$ mode is anti-resonant with the $ARE_{01}$ mode and remains confined to the core.

At certain values of $D/\lambda$, anti-crossings appear between the $LP_{01}$ mode and the $q$-th order transverse mode in the glass walls of the AREs, following the simple relationship [19]:

$$\left(\frac{D}{\lambda}\right)_q \approx \frac{q}{2(t/D)\sqrt{n_g^2 - 1}} \ . \tag{2}$$

The vertical dashed lines in Fig. 2a indicate the first two of these resonances, for $t/D = 0.01$. In the vicinity of these points the $LP_{01}$ mode leaks rapidly through the resonant AREs into the solid glass sheath, yielding loss values that are close to those of an isolated thick-walled dielectric capillary [17]; the result is a strong reduction in $FOM_{11}$ (see Fig. 2b). Away from these narrow regions, however, the $FOM_{11}$ remains relatively high – a consequence of the fact that the indices of the $LP_{11}$ and $ARE_{01}$ modes remain close to each other. The result is very strong $LP_{11}$ mode suppression over all the ranges of $LP_{01}$ mode transmission.

For values of $D/\lambda$ between 20 and 44 the $LP_{01}$ loss is below 0.3 dB/m. For larger values (away from resonances), however, it drops dramatically, reaching ~1 dB/km at $D/\lambda = 78$ and even lower values at $D/\lambda = 125$ (away from the glass wall resonances). Note however that at large $D/\lambda$ values the loss is likely to increase due to micro and macro-bending, which is not considered in the current simulations. Strong coupling to the $ARE_{01}$ mode results in very high $LP_{11}$ mode loss and $FOM_{11}$ values from ~1000 for $D/\lambda > 20$ to ~10,000 for $D/\lambda > 60$ – values that are higher than any previously reported in the literature.

In summary, the $LP_{01}$ mode transmission windows are limited by capillary wall resonances at shorter wavelengths (large $D/\lambda$) and by weak confinement at very long wavelengths (small $D/\lambda$). Broadband phase matching between $LP_{11}$ and $ARE_{01}$ modes (Fig. 2a) means that the hESM structure provides robust single-mode guidance at all wavelengths within the $LP_{01}$ transmission windows, the positions of which can be adjusted by varying $t/D$. We note that for large $D/\lambda$, higher order core modes are more likely to be excited when the fibre is subjected to stress because of the reduced index splitting compared to the $LP_{01}$ mode. Note, however, that the $LP_{11}$ mode will still have an index that is closer to the $ARE_{01}$ index, resulting in higher $LP_{11}$ loss.

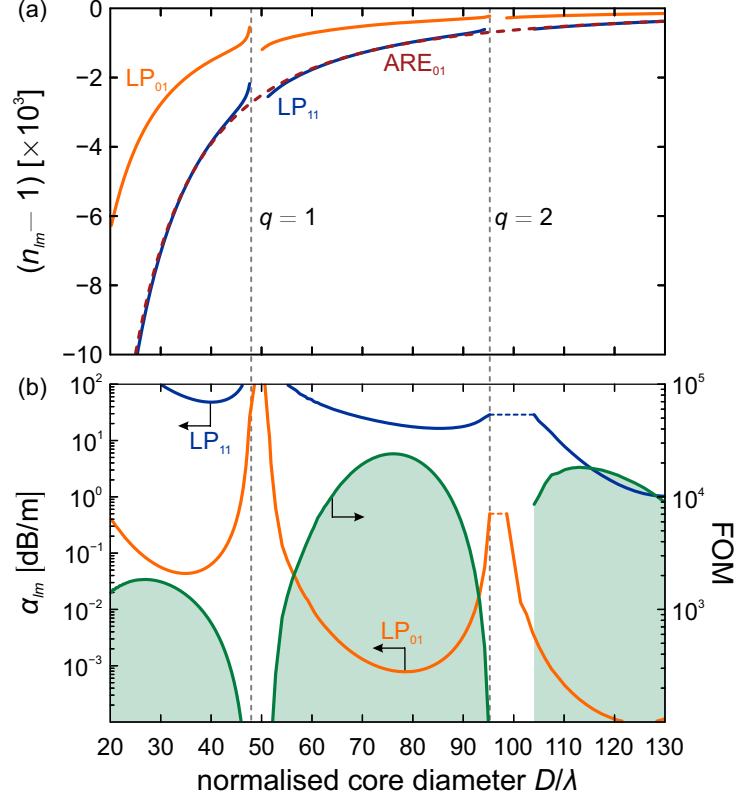

Fig. 2. (a) Numerically computed index $n_{lm}$ of the $LP_{lm}$ mode, plotted against $D/\lambda$. (b) Corresponding leakage loss (LH axis) and FOM (RH axis) at $d/D = 0.68$ and $t/D = 0.01$. The colour coding, labelling and additional structural parameters are the same as in Fig. 1. Note that around the $q = 2$ resonance the computational accuracy decreases because of strong modal complexity. The horizontal dashed lines in (b) indicate the limits of numerical accuracy.

We stress that Fig. 2b allows one to conveniently design an hESM PCF for operation in a desired wavelength range. For example, robust single-mode guidance ($FOM_{11} \sim 4{,}000$) with an $LP_{01}$ loss of ~40 dB/km could be achieved at 1 µm with a core diameter of 35 µm and a wall thickness of 350 nm – parameters that are easily accessible with current fabrication techniques.

On the other hand, broad-band transmission, e.g., for supercontinuum generation or spectroscopy, can be obtained by suitably adjusting $t/D$. For example, an hESM PCF with $D = 30$ µm and $t = 300$ nm would provide single-mode guidance with $LP_{01}$ loss less than 0.1 dB/m in the wavelength bands from ~0.7 to 1.2 µm and from ~0.3 to 0.6 µm (note that additional transmission bands exist in the deep ultraviolet).

## 3. Analytical model

To understand why maximum HOM suppression occurs at $d/D = 0.68$ for all wavelengths (except in the vicinity of ARE wall resonances, see Eq. (2)) we now present an analytical model in which the core and the AREs are treated as thick-walled capillaries (see Fig. 3a). The modal indices of the $LP_{lm}$ modes in a thick-walled capillary can be approximated by the modified Marcatili-Schmeltzer expression [17]:

$$n_{lm} = \sqrt{1 - \left(\frac{u_{lm}}{\pi f}\right)^2 \left(\frac{\lambda}{d_i}\right)^2} \tag{3}$$

where $u_{lm}$ is the $m$-th zero of the Bessel function $J_l$ and $d_i$ is the inner diameter of the capillary. The parameter $f$ is used to heuristically fit the analytical values from Eq. (3) to the results of FE simulations of the core and AREs. It corrects for the non-circular core and the finite wall thicknesses of the AREs.

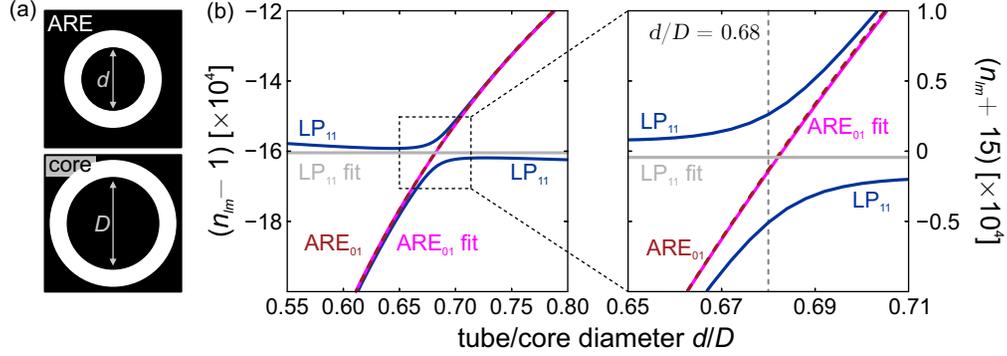

Fig. 3. (a) Thick-walled capillaries used to approximate the core and AREs in the analytical "toy" model. (b) Modal indices of the $LP_{01}$ and $ARE_{01}$ modes, calculated using Eq. (3) assuming isolated thick-walled capillaries in each case. Also plotted are the numerically evaluated indices of the hybrid $LP_{11}/ARE_{01}$ modes for the same parameters as Fig. 1a, i.e., $t/D = 0.01$ and $D/\lambda = 20$ (for this scaling the curves for the (fitted) $ARE_{01}$ modes are almost on top of each other). The right-hand panel is a zoom into the anti-crossing.

Fig. 3b plots the refractive index of the two $LP_{11}/ARE_{01}$ hybrid modes together with the fitted values for the $LP_{11}$ mode (zero line) computed using Eq. (2) with fit parameters $f_{co} = 1.077$ for the core and $f_{ARE} = 0.991$ for the ARE. Analytical and numerically computed values are in excellent agreement. The convenient analytical form of Eq. (2) allows one to derive a simple expression for the $d/D$ value at which the $LP_{11}$ and $ARE_{01}$ modes couple optimally:

$$\frac{d}{D} = \frac{u_{01}}{u_{11}} \frac{f_{co}}{f_{ARE}} = 0.682 \ . \tag{4}$$

This yields values that agree very well with those obtained from FE modelling (Fig. 1). If no correction is used, i.e., $f_{co} = f_{ARE} = 1$, the resulting value of $d/D$ is 0.628, i.e. only 8% less than the value obtained from FEM. Eq. (4) provides a convenient rule-of-thumb for designing robustly single-mode hESM PCFs. To a first approximation it depends neither on the refractive indices nor on the absolute physical dimensions of the fibre, making the design scalable. This means that, provided the ratio d/D is maintained, it becomes possible to design large-core hESM PCFs.

We note that Eqs. (3) and (4) can also be used to find structural parameters where other higher order modes (e.g. the $LP_{21}$ mode) are optimally suppressed. For example, for the almost degenerate $LP_{21}$ and $LP_{02}$ modes the value of $d/D$ obtained from Eq. (3) is very close to 0.53, which agrees with the optimum value obtained by FE modelling.

## 4. Conclusions and implications

A single ring of six non-touching thin-walled ARE capillaries, mounted inside a thick-walled glass capillary, provides robust single-mode $LP_{01}$ guidance at all wavelengths where the fibre guides with low loss. The broad-band single-mode behaviour of this hESM PCF is readily explained using an analytical model, resulting in the convenient condition $d/D = 0.68$ for a robustly single-mode hESM PCF, independent of the absolute size of the structure. Finite element modelling shows that the overlap integral of the $LP_{01}$ mode with the glass regions is

~0.003% at $D/\lambda = 25$, yielding a theoretical damage threshold of ~$10^{18}$ W/cm² (for fs pulses and silica glass) - a value similar to that seen in kagomé-PCF [4]. This makes hESM PCF highly suitable for ultrafast nonlinear optics in gases at extreme intensities. The design is scalable to large core diameters, making it suitable for ultra-high quality transmission of high power laser light, and it can be feasibly realized from many different glasses and polymers using state-of-the-art fibre drawing techniques. The ability to transmit a pure $LP_{01}$ mode at low loss levels is very important in many applications, for example, in-fibre chemical sensing, particle delivery, pulse compression and laser machining.